# Effects of substrate relaxation on adsorption in pores

Hye-Young Kim *, *Department of Chemistry and Physics, Southeastern Louisiana University, Hammond, Louisiana 70402*

Silvina M. Gatica, *Department of Physics and Astronomy, Howard University, 2355 Sixth Street NW, Washington, DC 20059*

George Stan, *Department of Chemistry, University of Cincinnati, Cincinnati, OH 45221-0172* and

Milton W. Cole, *Department of Physics, Penn State University, University Park, PA 16802*

* Corresponding author: Hye-young.Kim@selu.edu

Abstract

Fluids in porous media are commonly studied with analytical or simulation methods, usually assuming that the host medium is rigid. By evaluating the substrate's response (relaxation) to the presence of the fluid we assess the error inherent in that assumption. One application is a determination of the ground state of $^3$He in slit and cylindrical pores. With the relaxation, there results a much stronger cohesion than would be found for a rigid host. Similar increased binding effects of relaxation are found for classical fluids confined within slit pores or nanotube bundles.

The study of fluids in pores is important for both fundamental science (physics of reduced dimensionality) and applications (gas storage, purification, reactions and separations) [1-5]. Many calculations have been used to study these systems, often using simplified descriptions of the confining geometry (cylindrical or slit pore models). In many cases, an additional assumption is made- that the host material is rigid. In that approximation, the only role of the substrate is to provide a *static* potential energy function V(**r**) and a corresponding force, -∇V(**r**), on the adsorbed molecules. However, Newton's third law assures us that the molecules comprising the host material experience an equal, but opposite, force to that experienced by the adsorbate molecules. Is this reaction force important? In some cases, the answer is no; neglect of the substrate motion may be justified, for example, if its constituent molecules are massive or rigidly bound by strongly cohesive forces. This paper addresses several instances where the static substrate model does *not* adequately describe the fluid's behavior; indeed, the substrate's reaction to the fluid may lead to qualitative changes in the thermodynamics of adsorption. We consider a set of specific systems which manifest effects of pore relaxation; these are $^3$He (at temperature T=0) and classical fluids inside an individual nanotube (NT), a slit pore or a bundle of NTs. The extension to other systems is discussed briefly.

$^3$He in confining geometries is a particularly interesting system to explore because the existence of its bound states is marginal and therefore susceptible to perturbations. It is the only atomic system that does not condense in either one dimension (1D) or 2D. In contrast, $^4$He *is* bound as a 1D liquid, by just 2mK, the difference arising from the smaller zero-point energy (ZPE) of $^4$He [6]. In 2D, liquid $^4$He is relatively strongly bound, by ~0.9 K, while $^3$He does not condense because of fermi statistics and the higher ZPE [7]. In contrast, it has been shown that $^3$He inside a (rigid) carbon NT forms a weakly bound liquid, with cohesive energy ≤ 24 mK; the latter maximum value is found for a NT of optimized radius, R~ 0.7 nm [8]. This quasi-2D liquid state exists because the fluid density extends radially throughout the pore, reducing the effect of the interatomic, hard-core repulsion. (This condensed phase is an analogue of bound states of alkali gases in quasi-1D, which occur even in the absence of attractive pair potentials [9].) In the present study, instead, we consider a small radius NT, R<0.35 nm. The transverse displacements ($\mathbf{r}_\perp$) of the atoms are then limited to the vicinity of the NT (z) axis by the steep rise of the potential away from this axis.

Our first system involves a NT of equilibrium radius $R_0$ *prior to* the entry of any gas. We find the combination of radius R and 1D $^3$He gas density ρ=N/L which minimizes the ground state (gs) energy per atom E(ρ;R), in a tube of length L (L →∞). E(ρ;R) has contributions from the elastic energy of the tube, $E_{NT}(R)$, the atoms' mutual interactions and their interactions with the NT:

N E(ρ;R) = L $E_{NT}$(R) + N [$\varepsilon_{1D}$(ρ) + $E_{int}$(R)]          [1]

Here, $\varepsilon_{1D}$(ρ) is the energy per atom of 1D $^3$He, which has been evaluated with many-body techniques [6]. Since this function is minimum at ρ=0, the gs is a *gas*, as is the gs of quasi-1D $^3$He within a rigid nanotube of small radius. $E_{int}$(R) is the energy per atom due to $^3$He-NT interactions, including $\hbar\omega$, the transverse ZPE. The separability of longitudinal and transverse motions, implicit in Eq. 1, is a good approximation when $1/\rho \gg \langle r_\perp \rangle \sim [\hbar/(m\omega)]^{1/2}$, the rms transverse excursion from the axis [10]. This criterion is well-satisfied for $^3$He in small R tubes. The elastic energy is taken to be a quadratic function of deviations of the R from its initial equilibrium value $R_0$ (justified *a posteriori*):

$$E_{NT}(R) = E_0 + (k/2)(R-R_0)^2 \qquad [2]$$

We ignore the initial energy, $E_0$, henceforth, since it is a constant. For the interacting system's gs, we need to consider only uniform radial fluctuations, without azimuthal ($\phi$) fluctuations (Anisotropic "squash mode" fluctuations [11] couple to the adsorbate only in higher order). We expand the atom-NT coupling energy:

$$E_{int}(R) = E_{int}(R_0) + \phi(R-R_0) + (\lambda/2)(R-R_0)^2 \qquad [3]$$

Each term includes the potential energy on the tube axis and the transverse ZPE. The second term and the (usually less significant) third term in this equation express the fact that an atom's energy is a sensitive function of R. This dependence, involving expansion coefficients $\phi \equiv [\partial E_{int}/\partial R]_{R0}$ and $\lambda \equiv [\partial^2 E_{int}/\partial R^2]_{R0}$, provides the energy incentive for R to change from $R_0$. At fixed $\rho$, the gs equilibrium value of the radius, $R_{eq}$, is determined by minimizing Eq. 1 with respect to R. The resulting shift is

$$\Delta R \equiv R_{eq} - R_0 = -\rho\phi/[\rho\lambda + k] \qquad [4]$$

Since $[\rho\lambda+k]>0$, Eq. 4 shows that $\Delta R>0$, as expected, if $E_{int}(R)$ decreases with increasing radius ($\phi<0$); this is the case of a *very* small pore. The reverse is true in a larger radius pore. Having minimized with respect to R, the resulting energy per atom is a new function of density $\varepsilon_{new}(\rho)$:

$$E(\rho;R_{eq}) \equiv \varepsilon_{new}(\rho) \approx E_{int}(R_0) + \varepsilon_{1D}(\rho) - \rho\phi^2/[2(\rho\lambda+\kappa)] \qquad [5]$$

This energy is a constant, the original adsorption energy, plus the energy per atom of the interacting fluid, shifted downward by the *relaxation energy shift*, $\Delta\varepsilon(\rho)$, which is nearly proportional to $\rho$:

$$\Delta\varepsilon(\rho) \equiv \varepsilon_{new}(\rho) - \varepsilon_{1D}(\rho) - E_{int}(R_0) \approx -C(R_0)\rho \qquad [6]$$

$$C(R_0) = [\phi^2/(2(\rho\lambda+k))]_{R0} \qquad [7]$$

The coupling coefficient C(R) has been computed from the He/NT interaction [12] and values of the coefficient k derived from the NT radial breathing mode frequency, $\omega_{RBM}$ =A/R, where A~ 105 cm$^{-1}$.nm[13]. To find the gs of the coupled $^3$He/tube system, we find a minimum in the "new" energy function $\varepsilon_{new}(\rho)$. Using $\varepsilon_{1D}(\rho)$, provided by Krotscheck and Miller [6], Fig. 1 shows the resulting equilibrium densities $\rho_{eq}(R_0)$ as a function of $R_0$, for $R_0$ near the "optimal" radius $R_0$=0.305 nm, where $E_{int}(R_0)$ is a minimum. Also shown is the function $\varepsilon_{new}(\rho)$, evaluated at the density $\rho_{eq}(R_0)$. One observes that if the NT has the optimal radius, no relaxation occurs and the equilibrium $^3$He density is essentially zero, i.e., $^3$He is a gas. However, for slightly different values of $R_0$ the cohesive energy becomes substantial, ~ 1 K, once radial relaxation is included; the liquid density $\rho_{eq}$~ 1 to 3 nm$^{-1}$, a sensitive function of $R_0$. The basic point is this: the relaxation shift $\Delta\varepsilon(\rho)$ is roughly proportional to density, favoring a high density fluid. This effect may be thought of as a mean-field attraction mediated by NT contractions or expansions, analogous to the phonon coupling mechanism in superconductivity. A key difference is that this NT coupling is static, in the gs, analogous to the displaced oscillator shift in the energy of a particle coupled to a spring.

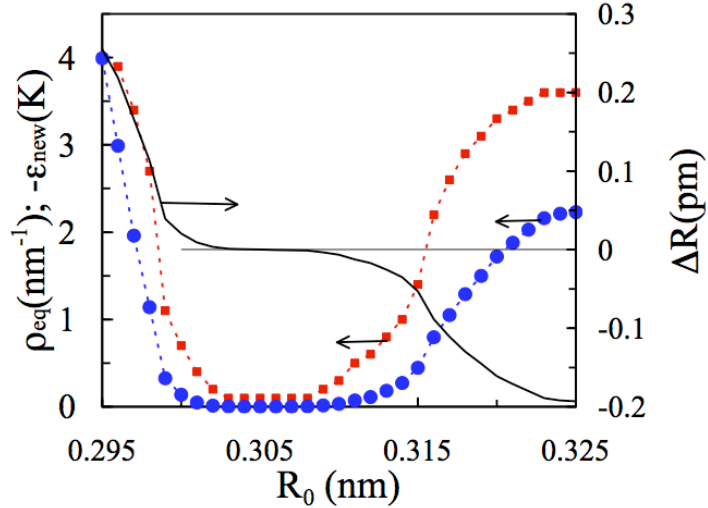

FIG. 1 (Color online). Equilibrium density $\rho_{eq}$ (red squares) and binding energy, $-\varepsilon_{new}(\rho)$ (blue circles); left scale, in units of nm$^{-1}$ and K, respectively, for $^3$He in tubes of initial radius $R_0$. On this scale, $\varepsilon_{1D}$ is negligible. The full curve shows $\Delta R=R_{eq}-R_0$ (right ordinate, in units of pm).

We note that $|\Delta R| < 0.00025$ nm, over the range of radii shown in Fig. 1, consistent with the harmonic expansion in Eq. 2. Qualitatively similar effects of NT relaxation are found for $^4$He, although that system is self-bound even without taking the relaxation into account. In either case, the presence of a finite equilibrium density implies quite different correlations than occur for a gas. From an experimental perspective, a key result is that low T thermal properties arise from density fluctuations (1D sound) about this gs density. This behavior will be explored in future work.

An analogous treatment is possible for $^3$He adsorbed within a slit pore, of initial width $L_0$. Omitting details, to be presented elsewhere, the resulting energy assumes a similar form to that in the NT case; the pore width changes from $L_0$ to an equilibrium value $L_{eq}$, when the fluid 2D density is $\theta$:

$$E_{total}(L_{eq}, \theta) \approx E_{int}(L_0) + \varepsilon_{2D}(\theta) - [\theta/(2k_{slit})] (\partial E_{int}/\partial L)^2 \qquad [8]$$

Since this expression has a minimum at finite $\theta$, the gs of $^3$He is a liquid, in contrast to the 2D gas found when the wall separation is fixed and small. A key parameter in this analysis is the coefficient $k_{slit}$, analogous to k in Eq. 2. Its value depends on the structure of the pore but it is, unfortunately, not known, in general. We may express it in terms of a dimensionless multiple ($\gamma$) of the force constant associated with the direct van der Waals (VDW) interaction between the two half-spaces bounding the pore: $k_{SLIT} \equiv -\gamma (\partial^2/\partial L^2) V_{slit}$. For our numerical work, the interaction $V_{slit}$ is chosen to be that corresponding to an empty graphitic pore of the specified width $L_0$.

Having addressed this extreme quantum fluid in two environments, we turn to the problem of a classical fluid. First, we consider the slit pore geometry, when a single layer film is sandwiched between two confining half-spaces. Thermodynamic perturbation theory yields the free energy change of this film due to its coupling to the flexible substrate, involving a coefficient $\Gamma$:

$$\Delta F(\theta)/N = -\Gamma \theta \qquad [9]$$

The corresponding change in the spreading pressure $P_{2D}$ is given by $\Delta P_{2D} = -\Gamma \theta^2$. The venerable, but approximate, VDW equation of state, with $\beta^{-1} = k_B T$, is

$$P_{2D} = \theta/[\beta(1-B\theta)] - a \theta^2 \qquad [10]$$

Here B=$\pi\sigma^2/2$ is the "hard-core" parameter for a gas with diameter σ. Thus, the perturbation corresponds to a change Δa = - Γ in the attractive contribution to the pressure. Since the 2D critical temperature $T_c$ is given in this theory by the relation $k_B T_c$ =8a/(27B), the change $\Delta T_c$ due to the relaxation satisfies $\Delta T_c/T_c = \Delta a/a = \Gamma/a$.

Fig. 2 shows results obtained for various gases in this geometry as a function of the reduced width, relative to $\sigma_{gs}$, the Lennard-Jones (LJ) length parameter for the gas-substrate atomic interaction. $\Delta T_c/T_c$ is large, except for values close to L=1.72 $\sigma_{gs}$, for which there is no effect because that is where the gas-surface interaction energy of the system is minimum.

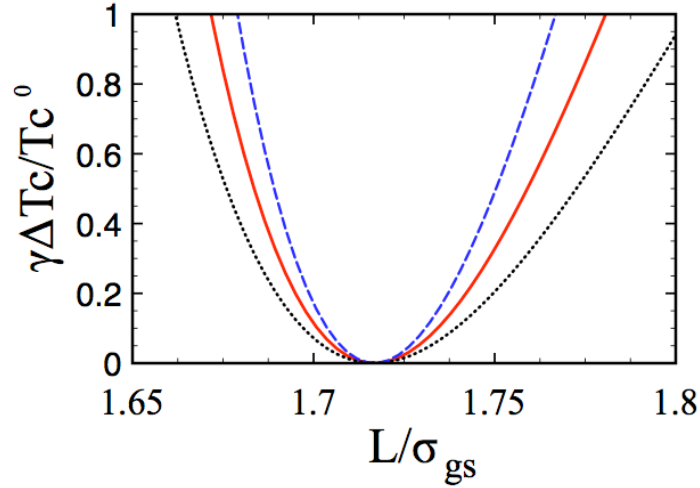

FIG. 2 (Color online). Relative change in $T_c$ in a slit pore of intial separation L. The gases shown are He (short-dashed black), Ar (solid red) and Xe (long-dashed blue).

We consider next the case of a classical fluid within a small cylindrical pore, adapting (to Ar near T=30 K) the cylindrical model applied above to $^3$He at T=0. Using thermodynamic perturbation theory, we evaluate the change in free energy due to radial compression, or expansion, from which we compute the change in chemical potential, $\Delta\mu(\rho,T) = -C_{pore}\,\rho$. Then $\mu(\rho,T) = \mu_0(\rho,T) - C_{pore}\,\rho$. The function $\mu_0$ has been computed previously by our group [14] for a LJ fluid, using the 1D equation of state derived by

Gürsey [15]. Fig. 3 shows results for this function, with and without tube relaxation. Only in the former case is the curve nonmonotonic.

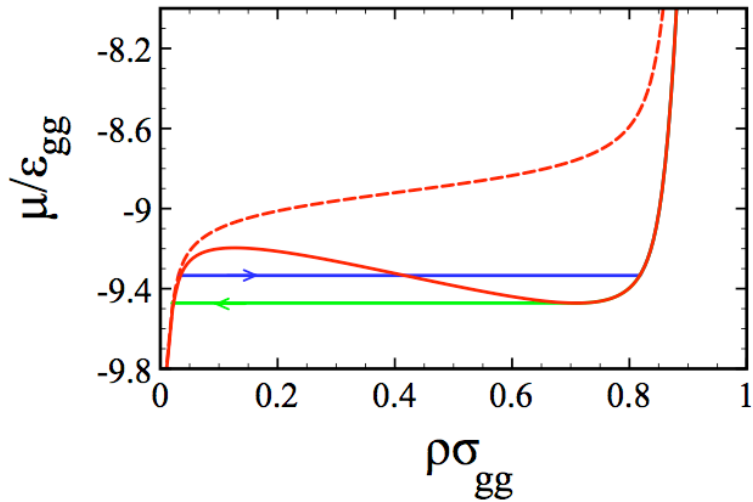

FIG. 3 (Color online). $\mu$ as a function of $\rho$ for Ar at 30 K within a NT of radius 0.325 nm (functions reduced by the LJ parameters of Ar). Dashed curve is for a rigid tube; full curve includes tube expansion. Hysteretic imbibition shown schematically: filling (right-pointing arrow) at higher $\mu$, then emptying (left-pointing arrow).

We speculate that the behavior in this figure may explain experimental data for metal-organic framework materials [16]. These isotherms show particularly large hysteresis loops. The arrows in the figure indicate our interpretation. At low $\mu$ (low pressure), the isotherm follows the low coverage contour, the system not recognizing that there exists a lower free energy state involving a contracted pore, occupied by fluid. Eventually, the barrier seen in the function $\mu(\rho,T)$ is surmounted, when the fluid enters the nearly empty, unrelaxed pore. The desorption branch of the isotherm follows the indicated path down to the minimum of the function $\mu(\rho,T)$, before emptying abruptly by switching to the low $\rho$ path. A serious analysis requires a treatment of the metastable state, a challenge in the venerable problem of capillary condensation [17], which is more complicated here because of the relaxation of the host material.

Finally, we discuss the problem of a classical gas adsorbed in the interior of NTs comprising a NT bundle [18]. ($^3$He or $^4$He are analogous) Our procedure is adapted from that used for an ensemble of "peapods", a collection of $C_{60}$ molecules within a nanotube bundle [13]. The free energy ($F_0$) includes a coupling between molecules in nearby parallel tubes, separated by b: $F_0 = NF_{1d} - \alpha N \rho$. Here, $F_{1D}$ is the free energy of a free-standing, unrelaxed 1d system and $\alpha=(9\pi/8) C_6/b^5$ is a geometrical coefficient proportional to the VDW interaction between atoms ($\sim C_6 r^{-6}$). To assess the substrate relaxation effects, we add the energetic contributions from the bundle variables (R, b) that are affected by the adsorbate: $F = F_0 + LF_{NT}(b,R) + NF_{int}(b,R)$. Expanding about the empty parameters ($b_0, R_0$), with shifts $\delta b$ and $\delta R$, one obtains a shift in pressure to $P = P_{1d} - (\alpha+K)\rho^2$. Numerical results for the ratio of $K/\alpha$, to be reported elsewhere, are similar to those in Fig. 2, except that the minimum occurs for the optimal radius, $R/\sigma_{gs} \sim 1.09$. The calculation of the critical temperature $T_c$ is analogous to that applied to the peapod problem [14], *i.e.,* the anisotropic Ising model. The resulting $T_c$ values are even higher than those found without relaxation.

In summary, we have considered problems in which relaxation of a pore plays a significant role in the thermodynamics of adsorption. The systems were chosen for computational simplicity, but the phenomenon is general. For a pore of *arbitrary* radius or shape, the effect can be evaluated by perturbation theory, starting from a Monte Carlo evaluation of the rigid pore case, from which the pressure on the wall can be computed. That pressure is the finite T analogue of the term $\phi \to -(\partial F/\partial R)$ in Eq. 3. The wall's response and the corresponding free energy change are computed, as in Eq. 5. This perturbative procedure can be iterated if the change in pore shape is not small.

We note that the "correction" described here becomes particularly large when a transition, such as capillary condensation, occurs. The system is particularly susceptible to any "environmental" change then; the response function (film compressibility) $\partial N/\partial \mu$ diverges at the capillary condensation transition. We note also that adsorption data on activated carbons often yield "unphysical" gaps in the pore width distribution derived from inverting the data [19]. Such gaps can arise when the host pores relax to avoid energetically costly adsorption in pores of these widths.

This research was supported by NSF DMR-0505160. We are indebted to Angela Lueking, Dave Weiss and Karl Johnson for stimulating comments and to Eckhard Krotscheck for performing calculations of $\varepsilon_{1D}(\rho)$ for 3He used in this paper.